\documentclass{ws-procs975x65}

\usepackage{graphicx}
\usepackage{bm}
\usepackage{epsf}

\def\agt{\mathrel{\raise.3ex\hbox{$>$}\mkern-14mu\lower0.6ex\hbox{$\sim$}}}
\def\alt{\mathrel{\raise.3ex\hbox{$<$}\mkern-14mu\lower0.6ex\hbox{$\sim$}}}
\newcommand{\beq}{\begin{equation}}
\newcommand{\eeq}{\end{equation}}
\newcommand{\beqn}{\begin{eqnarray}}
\newcommand{\eeqn}{\end{eqnarray}}

\begin{document}



\title{MAGNETIZED HYPERMASSIVE NEUTRON STAR COLLAPSE: \\
a candidate central engine for short-hard GRBs}

\author{Branson C.\ Stephens}

\author{Matthew D.\ Duez\footnote{Current address:  
Center for Radiophysics and Space Research,
Cornell, Ithaca, NY 14853}}

\author{Yuk Tung Liu}

\author{Stuart L.\ Shapiro\footnote{Also at the 
Department of Astronomy and NCSA, University
of Illinois, Urbana, IL 61801}}

\address{Department of Physics, University of Illinois at Urbana-Champaign,
Urbana, IL 61801, USA}

\author{Masaru Shibata}

\address{Graduate School of Arts and Sciences, 
University of Tokyo, Komaba, Meguro, Tokyo 153-8902, Japan}


\begin{abstract}
Hypermassive neutron stars (HMNSs) are equilibrium configurations 
supported against collapse by rapid differential rotation and likely 
form as transient remnants of binary neutron star mergers. Though 
HMNSs are dynamically stable, secular effects such as viscosity or 
magnetic fields tend to bring HMNSs into uniform rotation and thus 
lead to collapse. We simulate the evolution of magnetized HMNSs 
in axisymmetry using
codes which solve the Einstein-Maxwell-MHD system of equations. 
We find that magnetic braking and the 
magnetorotational instability (MRI) both contribute to the eventual collapse 
of HMNSs to rotating black holes surrounded by massive, hot accretion 
tori and collimated magnetic fields. Such hot tori radiate strongly in
 neutrinos, and the resulting neutrino-antineutrino annihilation could 
power short-hard GRBs. 
\end{abstract}

\bodymatter

\section{Introduction and Methods}

Short-hard gamma-ray bursts (SGRBs) emit large amounts of energy 
in gamma rays~\cite{GRB} with durations $\sim 10^{-3}$--2 s and
may originate from binary neutron star mergers~\cite{GRB,GRB-BNS}. 
If the total mass of the binary is below a certain
threshold, a hypermassive neutron star (HMNS) likely forms as a 
transient merger remnant.~\cite{STU,STU2}  HMNSs have masses larger than the maximum allowed 
mass for rigidly rotating neutron stars and are 
supported against collapse mainly by rapid, differential rotation~\cite{BSS}. 

\begin{figure*}[t]
\vspace{-1cm}
\begin{center}
\epsfxsize=1.8in
\leavevmode
\epsffile{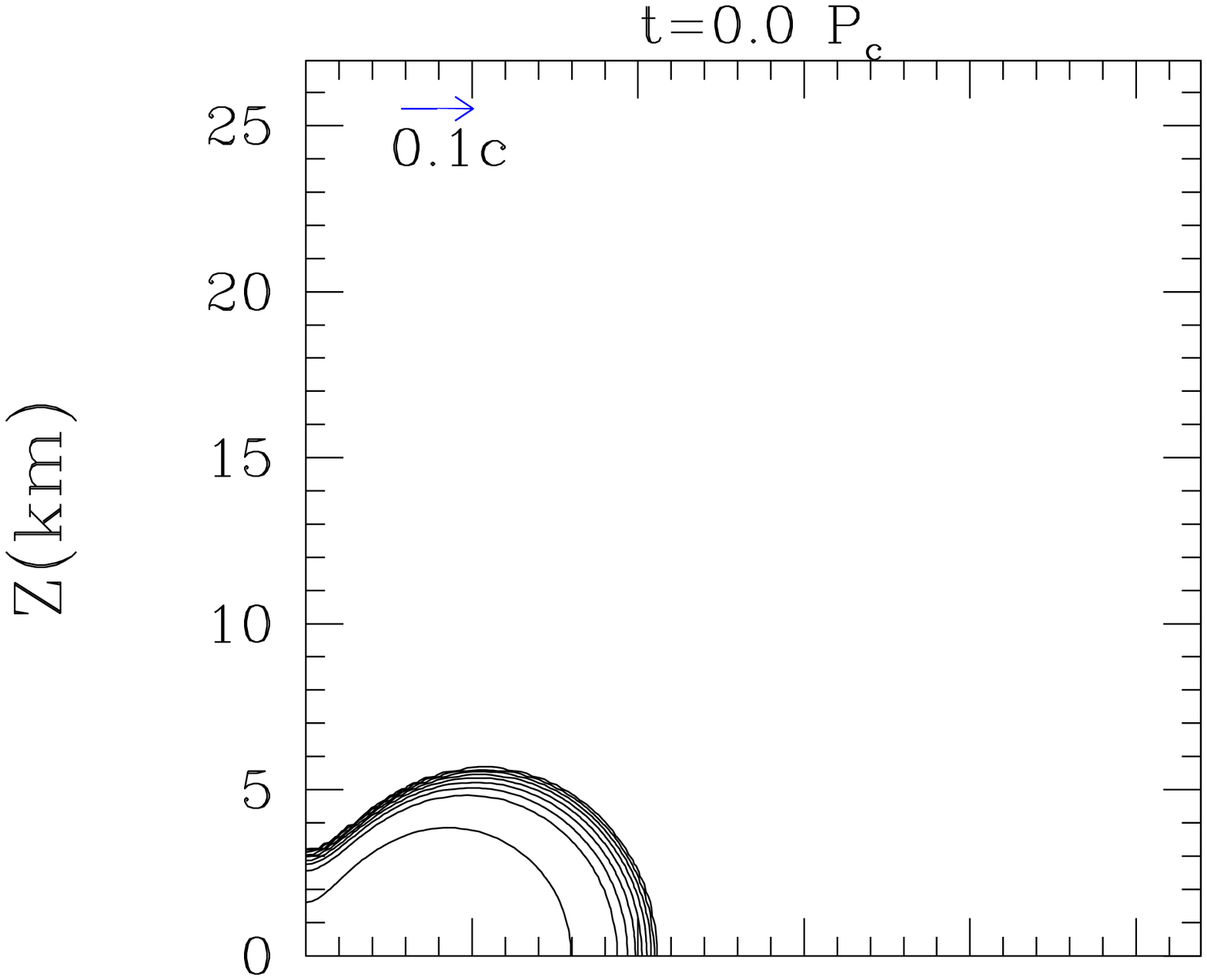}
\epsfxsize=1.8in
\leavevmode
\hspace{-1.55cm}\epsffile{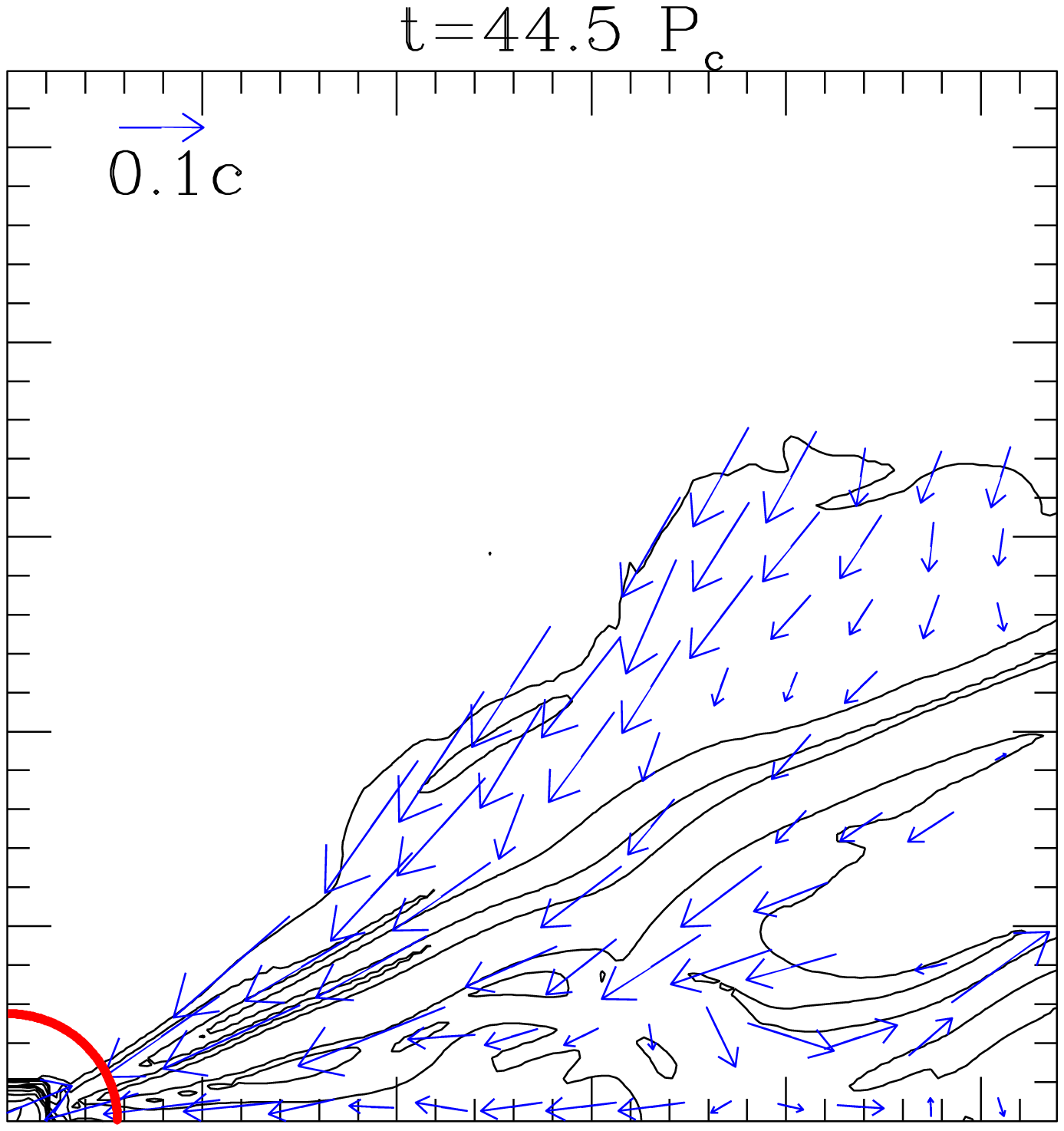}\\
\vspace{-1.5cm}
\epsfxsize=1.8in
\leavevmode
~\epsffile{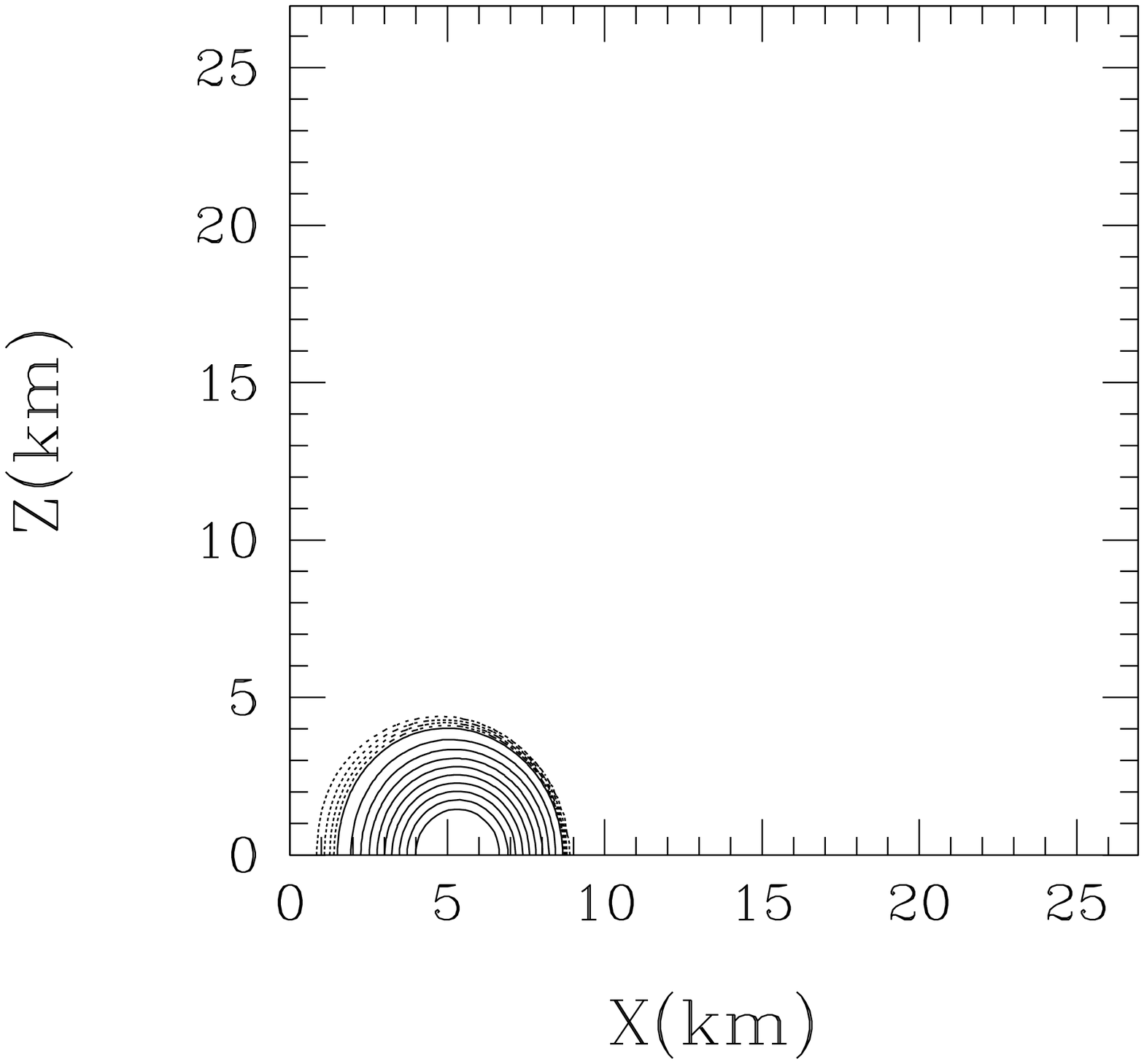}
\epsfxsize=1.8in
\leavevmode
\hspace{-1.55cm}\epsffile{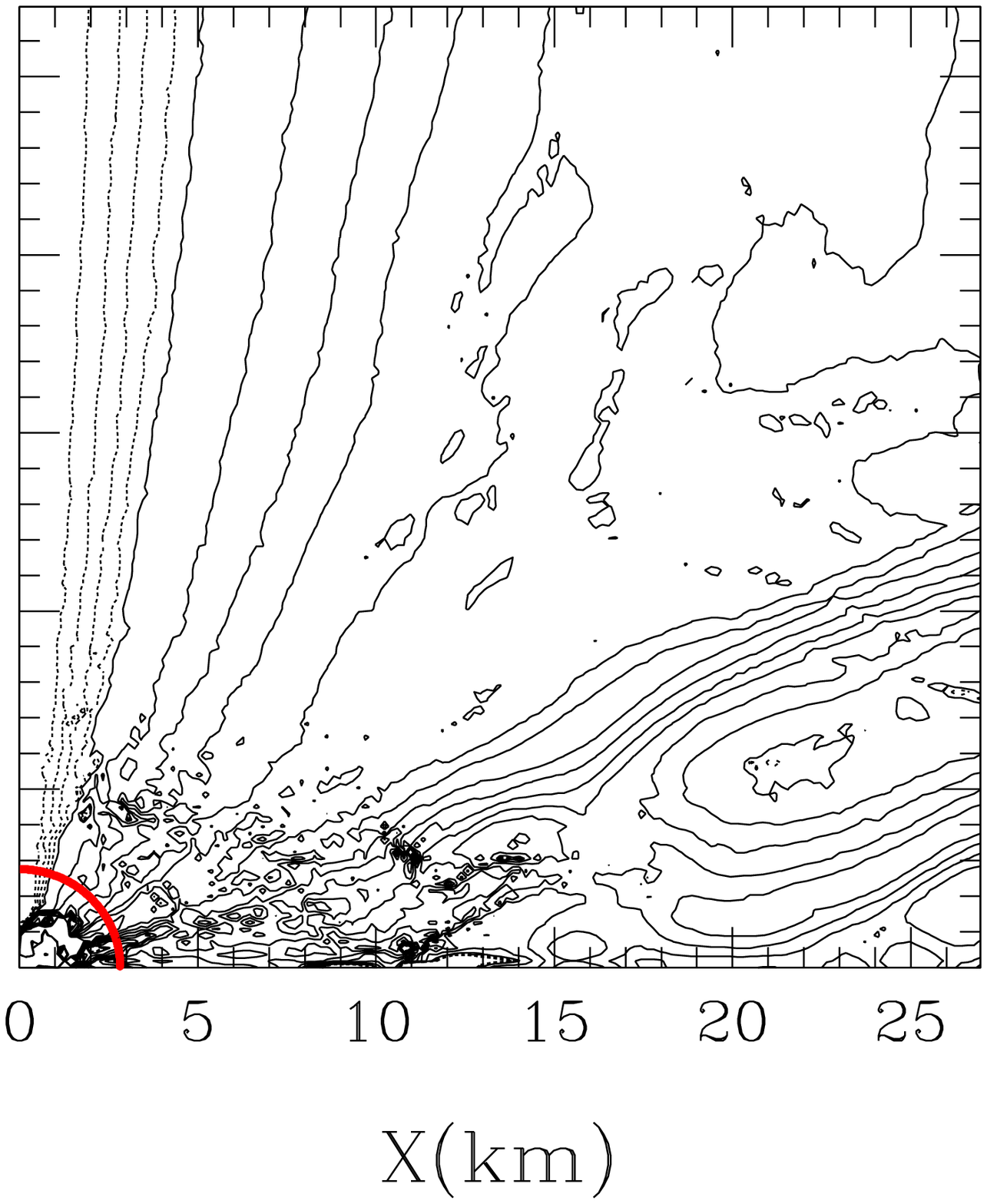}
\vspace{-6mm}
\caption{Upper panels: Density contours (solid curves) and 
velocity vectors at the initial time and at a late time. The contours are
drawn for $\rho=10^{15}~{\rm g/cm^3} \times 10^{-0.4 i}~{\rm g/cm^3}~(i=0$--9).
In the second panel, a curve with $\rho=10^{11}~{\rm g/cm^3}$
is also drawn.  The (red) circle in the lower left of the second panel
denotes an apparent horizon.  
The lower panels show the poloidal magnetic field lines
at the same times as the upper panels. 
\label{FIG1}}
\vspace{-6mm}
\end{center}
\end{figure*}

We have performed general relativistic magnetohydrodynamic (GRMHD) 
simulations of differentially rotating $\Gamma$-law HMNSs~\cite{DLSSS,grbpaper,bigpaper} 
using two new GRMHD codes~\cite{DLSS2,SS}.  Here, we consider an 
HMNS collapse model with a more realistic 
hybrid EOS.  For details of the EOS, the chosen HMNS model,
and the simulation,
see Shibata et al.~\cite{grbpaper} and Duez et al.~\cite{bigpaper} We construct a differentially
rotating HMNS with mass $M=2.65M_{\odot}$ and angular momentum $J=0.82GM^2/c$.
This HMNS is similar to one found in a BNS merger simulation~\cite{STU2}.
A small seed poloidal magnetic field is introduced into the HMNS 
as described in Shibata et al.~\cite{grbpaper}. 
These calculations are the first in general relativity to 
self-consistently generate a 
candidate GRB central engine (i.e., a rotating BH surrounded by a magnetized 
torus) from non-singular initial data.

\section{Results and Discussion}

In Figure~1, we show snapshots of the meridional density contours,
velocity vectors, and poloidal magnetic field lines at the initial time 
and at a late time.  The differential rotation of the HMNS winds
up a toroidal magnetic field, which then begins to 
transport angular momentum from the inner to the outer regions of 
the star (magnetic braking), inducing quasistationary contraction 
of the HMNS~\cite{DLSSS}.  After the toroidal field growth saturates, 
the evolution is dominated by the MRI~\cite{MRI}, which leads to turbulence,
thus contributing to the angular momentum transport.  The star eventually
collapses to a BH, while
material with high enough specific angular momentum remains in an 
accretion torus. 
The accretion rate $\dot M$ gradually decreases and eventually settles 
down to $\dot M \sim 10M_{\odot}/$s, giving an 
accretion timescale of $M_{\rm torus}/\dot{M} \sim 10~{\rm ms}$. 

To explore the properties of the torus, we calculate the surface 
density $\Sigma$ and the typical  thermal energy per nucleon, $u$.  
We find $u\approx 94$~MeV/nucleon, or equivalently, 
$T \approx 1.1 \times 10^{12}$ K. 
Because of its high temperature, the torus radiates
strongly in thermal neutrinos~\cite{PWF,MPN}. However, the opacity
inside the torus is approximately $\kappa \sim 7\times 10^{-15}
T_{12}^2$~${\rm cm}^2$~${\rm g}^{-1}$.  The optical 
depth is then estimated as $\tau \sim \kappa \Sigma \sim 7200 \Sigma_{18} 
T_{12}^2$, so that the neutrinos are
effectively trapped~\cite{MPN}. Here, $T_{12} = T/10^{12}~{\rm K}$,
$\Sigma_{18} = \Sigma / 10^{18}~{\rm g~cm}^{-2}$.  This 
regime of accretion has been described as a neutrino-dominated
accretion flow (NDAF)~\cite{ndaf}.

We note that the properties of the torus are not specific to the 
chosen initial data.  For example, we consider the model labeled 
star~A in Duez et al.~\cite{bigpaper}.  This is also an
HMNS model which collapses to a BH surrounded by a hot 
accretion torus under the influence of magnetic fields.  However, 
star~A has simple $\Gamma$-law EOS (not a hybrid EOS) and has 
a different rotation profile and compactness.  In this case, we 
find that the disk has $u\approx 5$~MeV/nucleon, which gives an 
optical depth of about 70.  Thus, the evolution of star~A also
produces a hyperaccreting NDAF.  

Returning to the hybrid EOS HMNS model, we estimate the 
neutrino luminosity in the optically-thick diffusion limit.~\cite{ST} 
We obtain $L_{\nu} \sim 2\times 10^{53}~{\rm erg/s}
(R/10~{\rm km})^2 T_{12}^2 \Sigma_{18}^{-1}$, which is comparable to
the neutrino Eddington luminosity~\cite{MPN}.  A model for the neutrino
emission in a similar flow environment with comparable $L_{\nu}$ gives
for the luminosity due to $\nu\bar\nu$ annihilation  
$L_{\nu\bar\nu}\sim 10^{50}~{\rm ergs/s}$~\cite{MPN}. 
Since the lifetime of the torus is $\sim 10$ ms, the total energy, 
$E_{\nu\bar\nu} \sim 10^{48}~{\rm ergs}$, may be sufficient to power
SGRBs as long as the emission is somewhat beamed~\cite{AJM}.  
Our numerical results, combined with accretion and jet 
models~\cite{MPN,AJM}, thus suggest that magnetized HMNS collapse 
is a promising candidate for the central engine of SGRBs. 

{\em Acknowledgments:}  
Numerical computations were performed at NAOJ, ISAS, 
and NCSA.  
This work was supported in part by Japanese Monbukagakusho
Grants (Nos.\ 17030004 and 17540232) and NSF Grants PHY-0205155 and
PHY-0345151, NASA Grants NNG04GK54G and NNG046N90H at UIUC.

\end{document}